\newcommand{\la}{\langle}
\newcommand{\ra}{\rangle}
\newcommand{\lla}{\la\!\la}
\newcommand{\rra}{\ra\!\ra}
\newcommand{\beq}{\begin{eqnarray}}
\newcommand{\eeq}{\end{eqnarray}}

\renewcommand{\theequation}{\thesection.\arabic{equation}}

\newcommand{\Sigpi}{\Sigma _{\pi N}}

\newcommand{\cl}{\centerline}

\newcommand{\btem}{\bibitem}
\newcommand{\TH}{T.\ Hatsuda}
\newcommand{\TK}{T.\ Kunihiro}
\newcommand{\PL}{Phys.\ Lett.\ {\bf B}}
\newcommand{\PTP}{Prog.\ Theor.\ Phys.}
\newcommand{\PR}{Phys.\ Rev.}
\newcommand{\PRL}{Phys.\ Rev. \ Lett.}

\newcommand{\HK}{T. Hatsuda and T. Kunihiro}
\newcommand{\KH}{T. Kunihiro and T. Hatsuda}

\documentstyle[12pt]{article}

\begin{document}

\hfill  RYUTHP 95-1

\hfill February 1995

\bigskip

\begin{center}

{\Large {\bf Chiral Symmetry and  Scalar Meson}} \\
{\Large {\bf in Hadron and Nuclear Physics}}
\footnote{
 Invited talk presented at YITP workshop on
   ``From Hadronic Matter to Quark Matter:  Evolution of Hadronic
Matter''.
Originally, the talk was entitled ''Hadron Properties and Chiral
Transition
in an Effective Theory'' by the organizers of the workshop.
To be  published in Prog. Theor. Phys. Supplement (Kyoto).
}

\vspace{1.0cm}

Teiji Kunihiro\\

\bigskip
Faculty of Science and Technology, Ryukoku University,
 Seta, Otsu-city\ \ \ \ \ \\
 520-21, Japan\\
\end{center}

\vspace{1.2cm}

\abstract

After giving  a short introduction to  the Nambu-Jona-Lasinio model
 with an anomaly term, we show the importance of the scalar-scalar
 correlation in the low-energy hadron dynamics, which correlation may be
 summarized by a scalar-isoscalar meson, the sigma meson.
The discussion is based on the  chiral quark model with the sigma-meson
 degrees of freedom. Possible experiments are proposed to produce the
elusive meson in a nucleus and detect it.
In relation to a precursory soft mode for the chiral transition,
the reason  is  clarified why the dynamic
 properties of the superconductor may be described by the
 diffusive time-dependent Ginzburg-Landau (TDGL) equation.
We indicate the chiral symmetry plays a significant role
 also in nuclei; one may say that
the  stability  of nuclei is   due to the chiral symmetry of QCD.

\newpage

\section{Introduction}

The salient features of  low-energy hadron physics or QCD
 may be summarized as follows;
(i) {\em Confinement of the colored quarks and gluons},\
(ii) {\em the  dynamical breaking of chiral symmetry} (DBCS),\
(iii) {\em $U_A(1)$ anomaly},\
(iv) {\em Explicit $SU_V (3)$ breaking},\
(v) {\em Success of the constituent quark model},\
(vi) {\em OZI rule and its  violation  in mesons and baryons}.
Hatsuda and the present author \cite{HK94} showed that
 a semi-phenomenological but unified description of the above facts
 is possible, emphasizing a dominant role of chiral symmetry in
 low-energy hadron dynamics. The description is based on the $SU(3)$
Nambu-Jona-Laisnio
 (NJL) model\cite{NJL}.
 The model embodies  three basic  ingredients of QCD, i.e.,
 the dynamical breaking of chiral symmetry (DBCS), $U_A(1)$ anomaly and
the explicit symmetry breaking
 due to the current quark masses. Various empirical aspects of QCD were
 shown  to be realized through  interplay among the three ingredients.
  It was emphasized that the constituent quark model and the
 chiral symmetry is reconciled in a chiral quark model. It was
 also shown that the chiral quark model can account for most of
  the empirical facts on baryons as well as the low-lying mesons.
 Furthermore,
 the NJL model allows us to study the change of hadron
properties in hot/dense medium in a self-consistent way.

In this report, we shall first give a short introduction of the
 NJL model with the anomaly term in section 2,
 which is the basis for the proceeding
 discussions. Then we pick up and rearrange  some topics dealt
in the above review article to clarify  significance of
  the sigma meson in the low-energy hadron dynamics.
  The sigma meson is a scalar and iso-scalar meson and
 the chiral partner of the pion; see section 3.
We show that  a chiral quark model incorporating the collectiveness
 in the scalar channel  summarized by the sigma meson nicely accounts for
  some properties of  lowlying baryons,
 such as  $\pi$-N $\Sigma$ term\cite{JK}. We also mention that the
 convergence radius of the chiral perturbation theory is related with
 the sigma meson mass.
In section 4, possible experiments are proposed to detect the elusive
 meson in the laboratory.

Next, we move to  the problem of  the fluctuation effects in hot QCD.
 This is also a topics dealt in Ref. \cite{HK94}, where  it is shown that  a
 precursory soft mode exist in the high temperature phase  for the chiral
transition in QCD.
 We shall  make a comment on the
 nature of the corresponding mode in the weak-coupling supercondoctor as
 described by  the BCS model\cite{BCS}: It will be clarified why the dynamic
 properties of the superconductor may be usually described by the
 diffusive time-dependent Ginzburg-Landau (TDGL) model; see section 5.

Finally, we indicate that the chiral symmetry plays a significant role
 in nuclei as well as in one hadron systems.
 Notice that nuclei are only stable bound systems
in the hadron world. One may say that this stability  is
 due to the chiral symmetry of QCD; see section 6.

%\newpage
\setcounter{section}{1}
\section{Brief summary of the $SU(3)$ NJL model with an anoamly term}
\setcounter{equation}{0}
\renewcommand{\theequation}{\arabic{section}.\arabic{equation}}

 Our model Lagrangian is the  generalized Nambu-Jona-Lasinio (NJL) model
with the anomaly term\cite{HK3a,HK3c,HK3f};
\beq
{\cal L} &=& \bar{q}i\gamma \cdot \partial q
+ \sum^{8}_{a=0}
{g_{_S} \over 2}[(\bar{q}\lambda_a q)^2 + (\bar{q}i\lambda_a
\gamma_5q)^2] -\bar {q}{\bf m}q+  g_{_D} [{\rm det}\bar{q}_i (1-\gamma_5) q_j
+h.c.],\nonumber \\
 \ \ \ &\equiv& {\cal L}_0  + {\cal L}_{SB} +{\cal L}_S + {\cal L}_{_{D}}\ \,
\eeq
where the quark field $q_i$ has three colors ($N_c=3$) and three flavors
($N_f=3$), $\lambda^a$ ($a$=0$\sim$8) are the Gell-Mann matrices with
$\lambda_0$=$\sqrt{2 \over 3}\bf{1}$. Here  ${\cal L}_0 +{\cal L}_S \equiv
{\cal L}_{NJL}$
 is the $U(3)$ generalization of the NJL model and has
 manifest flavor-$U_{_L}(3) \otimes U_{_R}(3)$ invariance.
 ${\cal L}_{_{SB}}$ is the explicit $SU_V(3)$ breaking part with
 $m_i$ ($i$=$u,d,s$) being the current quark mass.
Finally,
${\cal L}_{_{D}}$ in (2.1) denotes a term which
has $SU_{_L}(3) \otimes SU_{_R}(3)$  invariance but breaks the  $U_A(1)$
symmetry\cite{KOMAS,THOOF};
this term  is a reflection of the axial anomaly in QCD.
While ${\cal L}_S$ does not cause the flavor mixing, the anomaly term
does;
with the dynamical breaking of chiral symmetry,
  it induces effective 4-fermion vertices such as
$ \la\bar{d}d\ra (\bar{u}u)(\bar{s}s)$ and
$-\la\bar{d}d\ra (\bar{u}i \gamma_5 u)(\bar{s}i \gamma_5 s)$, where
 the former (latter) gives rise to
 a flavor mixing in the scalar (pseudo-scalar) channels.

For  notational convenience, we here  introduce  the bosonic
 variables $\Phi_{ij} = \bar{q}_j(1-\gamma_5)q_i$:
Note that $\bar {q}(1-\gamma _5)\lambda _aq={\rm Tr}[\lambda _a \Phi]$.
 The fact that ${\cal L}_{_{D}}$ represents the  $U_{_A}(1)$ anomaly can be
seen in the anomalous divergence of the flavor singlet axial current
$J_5^{\mu} = \bar{u}\gamma^{\mu}\gamma_5u+\bar{d}\gamma^{\mu}\gamma_5d+
\bar{s}\gamma^{\mu}\gamma_5s,$
\beq
\partial_{\mu}J^{\mu}_5  =  -4N_f g_{_D} {\rm Im}({\rm det}\Phi)
+ 2i\bar{q}m\gamma_5 q ,
\eeq
which is to  be compared with the usual anomaly equation written
in  terms of the
topological charge density of the gluon field,
$\partial_{\mu} J^{\mu}_5= 2N_f g^2 / {32\pi^2} F_{\mu \nu}^a
\tilde{F}^{\mu \nu}_a+ 2i \bar{q}m \gamma_5 q.$
Thus one may say that the determinantal 6-fermion operator
$-2g_{_D} {\rm Im}({\rm det}\Phi)$ simulates the effect
 of the gluon operator
 ${g^2 \over {32\pi^2}} F_{\mu \nu}^a \tilde{F}^{\mu \nu}_a$
\cite{THOOFT86} in the quark sector.
For further comments on the model lagrangian and the details of the
 following discussions, see the review paper \cite{HK94} or the original
 papers cited in the beginning of this section.

 In the self-consistent mean field theory,
 the Lagrangian can be rewritten as follows;
\beq
{\cal L}  =  {\cal L}_{_{MFA}} + {\cal L}_{res},
\eeq
where
 \begin{eqnarray}
{\cal L}_{_{MFA}} &=& \bar{q}(i\gamma \cdot \partial -M)q
- g_{_S} {\rm Tr}(\phi^{\dagger}\phi)
-2g_{_D}({\rm det}\phi + {\rm c.c.}), \nonumber \\
{\cal L}_{res} & = & g_{_S}:{\rm Tr}(\Phi^{\dagger}\Phi): \nonumber \\
&   & +g_{_D}:[{\rm Tr}(\phi\Phi^2)
-{\rm Tr}(\phi\Phi){\rm Tr}\Phi
-{1 \over 2}{\rm Tr}\Phi^2{\rm Tr}\phi+{1 \over 2}
{\rm Tr}\phi({\rm Tr}\Phi)^2 + {\rm h.c.}]: \nonumber \\
&   & + g_{_D}:({\rm det}\Phi + {\rm h.c.}):.
\end{eqnarray}
 Here the Fock terms are omitted,
the normal ordering is taken with respect to the Fock vacuum of
${\cal L}_{MFA}$
 and  $\phi$ is a diagonal 3 $\times $ 3 matrix defined in terms of the
 quark condensates as
\beq
 \phi  =  \langle  \Phi \rangle_0
 \equiv {\rm diag.}(\la \bar{u}u \ra, \la \bar{d}d \ra ,
 \la \bar{s}s \ra )
\eeq
 The ``constituent quark mass matrix"
 $M$ = diag.($M_u,M_d,M_s$) is given in terms of the condensates
\beq
M_u =  m_u - 2g_s \la \bar{u}u\ra - 2g_{_D} \la \bar{d}d\ra \la\bar{s}s
 \ra,
\eeq
 $M_d$ and $M_s$ with the subscripts $u,d$ and $s$ being replaced in a
  cyclic
 way.  The quark  condensates are in turn given with $M_i$;
\beq
\la \bar{u}u\ra =
 -i N_c {\rm Tr} \int {d^4p \over (2\pi)^4}
{1 \over {\gamma \cdot p - M}}
 = -\frac{N_c}{\pi ^2} \int_0^{\Lambda} p^2 dp
{{M_i} \over {\sqrt{M_i^2 + p^2}}}.
\eeq
Here the three momentum cut-off $\Lambda$ is introduced,
 and $\la \bar{d}d\ra$ ($\la\bar{s}s\ra$) is obtained by the
replacement $M_u$$\rightarrow$ $M_d$
($M_u$$\rightarrow$$M_s$).

The $q$-$\bar{q}$ collective excitations above the condensed vacuum
 are formed due to  the residual interaction
${\cal L}_{res}$ in (2.4). The 4-fermion part of ${\cal L}_{res}$
 can be decomposed into
  physical channels;
\begin{eqnarray}
{\cal L}_{res}^{(4)}& = &{1 \over 2}[(G_{\delta}\sum_{a}^{1,2}
+G_{\kappa^{\pm}}\sum_{a}^{4,5}
+G_{\kappa^0}\sum_{a}^{6,7}):S_a^2:
+\sum_{a,b}^{0,3,8}:S_a G_{ab}^S S_b:] \nonumber \\
& + &{1 \over 2}[(G_{\pi}\sum_{a}^{1,2}
+G_{K^{\pm}}\sum_{a}^{4,5}
+G_{K^0}\sum_{a}^{6,7}):P_a^2:
+\sum_{a,b}^{0,3,8}:P_a G_{ab}^P P_b:] \ \ ,
\end{eqnarray}
where we have introduced the composite operators
$S_a = \bar{q}\lambda_a q\ \ \ \ {\rm and} \ \ \ \
 P_a = \bar{q} i \gamma_5 \lambda_a q$.
The coupling constants ($G$'s) in ${\cal L}_{res}^{(4)}$ are summarized
in Table 2.1 of Ref.\cite{HK94}.
 The suffix of the coupling constants shows the relevant
channel where the residual interaction is active.
We note that  not only  ${\cal L}_{_S}$ but also  ${\cal L}_{_{D}}$
contribute to the residual 4-fermion interaction.

In Table 1, we summarize  basic physical quantities
calculated in the NJL model together with the  corresponding empirical
values.
The numerical values are obtained with the following parameter set;
$\Lambda = 631.4 {\rm MeV},\ \   g_{_S}\Lambda^2=3.67,\ \
g_{_D}\Lambda^5=-9.29,
\ \   m_s=135.7{\rm MeV},$
where we have used a three-momentum cutoff scheme.
\footnote
{
The corresponding invariant
 cutoff   for the non-strange (strange) quarks reads
$\Lambda_4 = 2 \sqrt{M^2 + \Lambda^2} = 1430\, (1640)\,$ {\rm MeV}
 with the constituent quark mass $M=335 \, (527)$MeV.
}

The predicted constituent quark masses
\beq
M_u=M_d=335{\rm  MeV}\ \ \ {\rm and}\ \ \ M_s=527 {\rm  MeV},
\eeq
  are  consistent with
the phenomenological masses extracted from the baryon magnetic
moments \cite{CLOSE}.

 The non-perturbative part of the condensate
 in Table 1 is defined  by subtracting the perturbative contribution
from
the full condensate \cite{HK2e}
\beq
 \la \bar{q}q\ra^{NP} & \equiv &  \la : \bar{q}q : \ra
 =  \la \bar{q}q \ra - \la \bar{q}q\ra^{pert.}
= -i N_c {\rm tr} \int {d^4p \over (2\pi)^4} [{1 \over \gamma \cdot p - M}
                                    -{1 \over \gamma \cdot p - m}] \ \ .
\eeq
The absolute value of the condensate and the $SU_V(3)$ breaking ratio
$\la \bar{s}s \ra^{NP}/ \la \bar{u}u \ra^{NP}$ in the NJL model
  agree well with those deduced from the QCD sum rules \cite{RECENT}.
It should be emphasized that the quark condensates to be deduced
 from the QCD sum rules are the non-perturbative part as
 defined above, while those  calculated in the lattice simulations
 are total ones; the total condensate of the $s$ quark is larger
in the absolute value than that of the non-strange quarks in accordance
 with the lattice result\cite{barkai}.

\bigskip

\cl{
\begin{tabular}{|c|c|c|}   \hline
& Theory & Empirical values  \\ \hline \hline
$M_u$ ($M_s$) & 335 (527)  & 336 (540) MeV \\
$\la \bar{u}u\ra^{NP}$ & $-(245)^3$   & $-(225\pm 25)^3$ MeV$^3$ \\
$\la \bar{s}s\ra^{NP}$/$\la \bar{u}u\ra^{NP}$ & 0.78  &  $0.8\pm 0.1$ \\
$m_{\pi}$ ($m_{K}$) & 138$^*$ (496$^*$)  & 138 (496) MeV\\
$m_{\eta}$ ($m_{\eta'}$) & 487 (958$^*$) & 549 (958) MeV\\
$m_{\sigma}$ ($m_{\sigma'}$) & 668 (1348) & $\sim 700$ ($\sim 1400$) MeV \\
$\Gamma_{\sigma \rightarrow 2 \pi}$  & $\sim 900$  & $\sim$Re
$m_{\sigma}$\\ $f_{\pi}$ ($f_{K}$) & 93.0$^*$ (97.7) & 93 (113) MeV\\
$f_{\eta}$ ($f_{\eta'}$) & 94.3 (90.8) & 93$\pm$9 (83$\pm$7) MeV \\
$\theta_{\eta}$ ($\varphi_{\sigma}$) & $-$21$^{\circ}$ ($-$6.8$^{\circ}$)
& $\sim -20^{\circ}$ ($-$) \\
$G_{\pi q}$ ($G_{Kq}$) & 3.5 (3.6) & $\sim 3.5$ ($-$) \\
$G_{\pi N}$ ($G_{\sigma N}$) & 12.7 (7 $-$ 10) & 13.4 ($\sim$10.0) \\
$\Sigma_{\pi N}$  & 49 $\pm 7$    & $45 \pm 10$ MeV\\ \hline
\end{tabular}
}
\bigskip
\noindent
Table 1:  Comparison of the theoretical estimates and the
experimental/empirical values of the basic physical quantities.  *
 indicates the quantity used as input.

\bigskip\par
Comments on other quantities listed in the table are given in  \cite{HK94}.
As for the sigma meson, we shall say a lot in the following section.
 Table 1 tells us  that the $SU(3)$ NJL model  reproduces the
fundamental physical quantities  in the accuracy of O(10\%-15\%).

%\newpage
\setcounter{section}{2}
\section{Role of the sigma meson in the QCD phenomenology}
\setcounter{equation}{0}
\renewcommand{\theequation}{\arabic{section}.\arabic{equation}}

\subsection{The Sigma meson}

 The sigma meson is the chiral partner of the pion for the
$SU_L(2)\otimes SU_R(2)$ chiral symmetry: In the
$(1/2, 1/2)$-representation, the sigma field $\sigma $ constitutes the
 quartet together with the three pions. This is well represented in the
 linear sigma model.
   The order parameter of the
chiral transition of QCD is the scalar quark condensate
$\la \bar q q\ra\sim \sigma $, and the vacuum is determined as the sate
where the effective  potential ${\cal V}(\sigma)$ takes the minimum.
  Let us denote the minimum point by $\sigma _0$.
 Then the   particle representing the quantum fluctuation
 $\tilde {\sigma}\sim \la :(\bar q q)^2:\ra$ is the sigma meson.
($\sigma = \sigma _0 + \tilde{\sigma}$). In this sense, the sigma meson
is analogous to the Higgs particle in the Weinberg-Salam theory,
where the Higgs field is the order parameter, and the quantum
fluctuation of the field around the minimum point of the Higgs potential
or the effective potential is the Higgs particle in the present world.

In the NJL model, the chiral symmetry is realized linearly like the linear
sigma models, hence the appearance of the sigma meson is
inevitable:
 The sigma meson mass $m_{\sigma}$ is predicted
 to be  twice of the constituent quark mass in the chiral limit
 \cite{NJL,HK1f},  hence $m_{\sigma}\sim 700$ MeV
 as in  the ladder QCD\cite{scadron}.

There exists, however,  a controversy on the identification of the
  nonet scalar mesons in the particle zoo\cite{PDATA}, and some people
are skeptical even about the existence of the
sigma  meson with a rather low mass, say about 600 to 800 MeV
\footnote{
Our view about the identification of the scalar mesons is given in
 chapter 3 of ref.\cite{HK94}.
}.
The origin of the such skepticism may be related to the facts that the
decay of the sigma  meson to two pions gives
 rise to a huge width $\Gamma= 400 \sim 1000$ MeV of the sigma meson, and
 that a possible coupling with glue balls with $J^{PC}= 0^{++}$ make the
 situation obscure.

  Nevertheless some studies \cite{SCAD,kamin}
 on the phase shift analysis of the pi-pi scattering in the scalar
 channel which claims a pole of  the scattering matrix in the complex
energy plane   with the real part Re$m_{\sigma}\simeq 500$ MeV and the
imaginary part Im$m_{\sigma}\simeq 500$MeV. Fig. 1 shows how the phase
 shift is reproduced with such a broad resonance \cite{kamin}: Actually
 the analysis shows the existence three resonances, i.e., $f_0(500),\
 f_0(975)$ and $f_0(1400)$ in the terminology in the PDG\cite{PDATA}.
 There are two sets of
 masses ($M$) and
 widths ($\Gamma$) of resonances obtained to fit the data. One set reads
 (in MeV);
$(M,\ \Gamma)= (506\pm 10,\ 494\pm5),(973\pm2,\ 29\pm2),\
(1430\pm 5,\  145\pm25)$
 for  $f_0(500),\  f_0(975)$ and $f_0(1400)$, respectively.
Another set of masses and widths is not so different from the above values,
 and predicts a $f_0$ meson, i.e., the sigma meson with the mass 505$\pm 10$
 MeV and the width 497$\pm5$ MeV.

 There is  also a preliminary experimental result at KEK\cite{shimizu},
 which seems to show a
 bump around 600 MeV with a width $\sim 400$ MeV in the reaction
$\pi^{-}$p $\rightarrow $ n$\pi^{0}\pi^{0}$.  The 2 $\pi ^{0}$ are detected
by 4 $\gamma$'s.  This is a clever experiment
 in the sense that  by confining to the 2$\pi^{0}$ channel,
 one can reject the iso-vector
 channel where we would have a huge yield from the rho meson
\footnote{
 Experimental status about the scalar
 particles is summarized in an Appendix of ref.\cite{HK2c} as well as
in ref.\cite{HK94}.  See these references for other experiments about
 the sigma meson.}

\vspace{1cm}
\cl{\fbox{{\bf Fig.1}}}
\vspace{1cm}

If  such a  scalar meson with a low mass is identified, many
 experimental facts  which have been mysterious  can be nicely accounted
for  in a simple way: The correlation in the scalar
 channel as summarized by such a scalar meson can account for the
 $\Delta I= 1/2$ rule for  the decay process
 K$^{0} \rightarrow \pi^{+}\pi^{-}$ or
 $\pi^{0}\pi^{0}$ \cite{morozumi};  see Fig.2.
In the meson-theoretical model for the nuclear
 force,   a scalar meson exchange with the mass range  500$\sim $ 700 MeV
is  indispensable to fully account for the state-independent attraction in
the  intermediate range\cite{DURSO}.
  These facts indicate that the scalar-scalar
 correlation is important in the hadron dynamics. This is in a sense
 natural because the dynamics which is responsible for the correlations
 in the scalar channel is nothing but the one which drives the chiral
 symmetry breaking.

\vspace{1cm}

\cl{\fbox{{\bf Fig.2}}}
\vspace{1cm}

In this section, we show that the correlations in the scalar channel
  as possibly described as   the sigma meson are  also essential
 in reproducing some
 interesting observables such as the $\pi$-N sigma term $\Sigpi$\cite{JK};
\beq
\Sigpi = \hat {m}\la \bar{u} u + \bar {d} d\ra.
\eeq
The empirical value of $\Sigpi$ is reported to be $45\pm 10$ MeV\cite{LOC}.
 The chiral perturbation theory fails in reproducing the empirical
value unless a huge strangeness content of the proton is
assumed\cite{JK}.
 Our  discussion will be  based on a chiral quark model\cite{CHQM}:
We shall
 show that the chiral quark model based on the NJL model can account for  the
 $\pi$-N $\Sigma$ term (and other related quantities in the scalar
channel) by taking into account the correlations in the scalar channel
as summarized by the sigma meson\cite{HK3d,HK3f,HK3g}; see also \cite{HK3b}.

 We remark also the convergence radius of the chiral perturbation theory
\cite{CH,HK3e}  is
 linked with the mass of the scalar meson.  The following subsections are
 a rewrite of a part of chapter 3 and 4 of ref.\cite{HK94} with an
emphasis put on the role of the sigma meson.  See ref.\cite{HK94}
and references cited therein for more details.

\subsection{$\pi$-N $\Sigma$ term in a chiral  quark model}

The basic quantities underlying the following discussion are the
quark contents of hadrons $\la h\vert \bar {q}_i q_i\vert h\ra \equiv
 \la \bar{q}_iq_i\ra_h$ ($i= u, d, s, ...$).  Actually, it is more
adequate to call them the scalar charge of the hadron. They
 are interesting in the relation  with the problems
 of the $\pi$-N $\Sigma $ term $\Sigma _{\pi N}$,
 the degree of the  Okubo-Zweig-Iizuka (OZI) rule\cite{OKUBO},
 the anomalous charm production in high-energy hadron-hadron collisions
 and so on.\cite{HK94}
 In this report, we confine ourselves on the problem of the
 quark contents of low-lying baryons in the scalar channel, i.e.,
 $\la \bar{q}_iq_i\ra _B$.

Feyman-Hellman theorem tells us that once the baryon mass $M_B$ is
known as a function of the current quark masses $m_i$, the quark
content of the baryon is calculated as,
\beq
 \la \bar{q}_iq_i\ra _B= \frac{\partial M_B}{\partial m_i}.
\eeq
Then the problem is to know  how $M_B$ is dependent on
 $m_i$ ($i= u, d, s, ...$).  In this respect,
  the chiral quark model is useful, where the notion of the chiral symmetry
 and the constituent quark model are reconciled: In this model, the dependence
 of $M_B$ on $m_i$ is known through the constituent quark mass $M_{i}$
 which is identified with the sum of the current quark mass $m_i$ and
 the mass $M^D_i$ generated by the dynamical breaking of
 chiral symmetry(DBCS);
$M_{i}=M_{i}(m_u, m_d, m_s, ...)$. On the other hand, $M_B$ is given
 in terms of $M_i$'s in a constituent quark  model and
 hence the   the current quark masses $m_i$'s.

Our theory may be described by the following effective Lagrangian
\beq
{\cal L}^{\rm eff}= {\cal L}_{\chi{\rm SB}} +
 {\cal L}_{``OGE''} + {\cal L}_{\rm conf},
\eeq
where ${\cal L}_{\chi {\rm SB}}$ is the interaction responsible for DBCS
 and is identified with ${\cal L}_{NJL}$. ${\cal L}_{\chi {\rm SB}}$
 is first
 switched on and then by adopting a successful constituent quark model,
 the spin-spin interaction between the constituent quarks
${\cal L}_{``OGE''}$
 and the  confining force ${\cal L}_{\rm conf}$ are switched on.
Such a two-stage approach was first discussed by Goldman and
Haymaker\cite{GOHA}.
 When  the NJL model is adopted to describe DBCS,  the
 constituent quark masses $M_u, M_d, M_s$ are given as
the solution of the gap equation
 Eq.(2.6). As  a successful constituent quark model, we
 adopt the Isgur-Karl model  \cite{CQM}; the baryon mass formula then
 reads as follows,
\beq
{M_B} = M_0+\sum _{i}^{u,d,s}[M_i+{a \over {2M_i}}]  +
b\sum_{i<j}{{{\bf \sigma}_i \cdot {\bf \sigma}_j} \over { M_iM_j}} .
\eeq
 Here $M_0$  represents
the contributions of the
confinement potential and the short-range color-electric interaction,
which are independent of the constituent masses.
$a/2M_i$
denotes the kinetic energy of the confined quarks. The spin-spin term
with the coefficient $b$
is the color-magnetic interaction responsible for the octet-decuplet
mass splitting.\footnote{The spin-spin interaction may include
 a contribution from the instanton-induced quark-quark interaction
\cite{OKA}.  We do not identify the origin of the spin-spin interaction here.}
  Instead of taking the detailed form of the quark
 wave functions inside the baryons, we take  the known
 masses of proton(938), $\Delta (1232)$ and $\Omega(1672)$
to extract  the parameters $M_0$, $a$ and $b$. The result is
$
a = (175.2 {\rm MeV})^2,\ \ M_0=-56.4 {\rm MeV},\ \ b=(176.4{\rm MeV})^3,
$
which gives the  baryon masses in a excellent agreement (in MeV unit);
\beq
\Lambda (1115)=1114,\ \  \Sigma (1193)=1186,\ \
\Sigma ^{\ast }(1385)=1372,\\ \nonumber
\ \ \ \   \Xi (1320)=1332, \ \ \ \  \Xi ^{\ast }(1507)=1519.
\eeq

Now, taking the derivatives of the mass formula Eq.(3. 4), one immediately
 finds that the quark content $\la \bar {q}_i q_i\ra_B$ or the scalar charge
 is given in terms of the scalar charge of the constituent quark,
\beq
Q_{ji}\equiv \frac{\partial M_{j}}{\partial m_i}=
\la \bar{q}_iq_i\ra _{q_j},
\eeq
with $i, j = u,d, s,...$.  Therefore let us first calculate and analyze
$Q_{ji}$.

$Q_{ji}$ can be obtained by differentiating the gap equation Eq.(2. 6)
 with the quark condensate Eq.(2.7).
  Due to the dynamical chiral symmetry breaking,
 $M_{u,d,s}$ are  composed of  two pieces, i.e.,
  the current quark mass and the dynamical mass
\beq
M_{u,d,s} = m_{u,d, s} + M_{u,d,s}^D \ \ \  .
\eeq
We stress that $M^D_{u,d,s}$
   has a purely  non-perturbative origin
 and is a non-linear function of $m_{u,d,s}$.

Using the NJL model, one can go further and identify the effect of
$M_{u,d,s}^D$  in a concrete form: Noting that
\beq
\frac {d\la \bar {q}_i q_i \ra}{dM_i}= \Pi ^{S}_{i}(q^2=0),
\eeq
where $\Pi ^{S}_{i}(q^2)$ is the zero-th order polarization in the
scalar channel due to the i-quark (i$=$u, d, s),
 one has
\beq
{\bf Q}=\bigl[{\bf 1}+ {\bf V}^{\sigma }\cdot {\bf \Pi }^{S}(0)
\bigr]^{-1},
\eeq
where
\beq
{\bf V}^{\sigma} =  2\cdot \left( \begin{array}{ccc}
g_s & g_{_D}\la\bar{s}s\ra  & g_{_D}\la\bar{d}d\ra \\
g_{_D}\la\bar{s}s\ra  &  g_s & g_{_D}\la\bar{u}u\ra \\
g_{_D}\la\bar{d}d\ra  & g_{_D}\la\bar{u}u\ra  & g_s
\end{array} \right)\ \
\eeq
is  the  interaction Lagrangian in the scalar channel in the flavor basis;
\beq
{\cal L}^{\sigma }_{res}
=\sum _{i,j=u,d,s}:\bar{q}_iq_i V_{ij}^{\sigma}\bar{q}_j q_ j:.
\eeq
  The off-diagonal terms of $Q_{ij}$ are
 proportional to $g_{_D}$, which  shows that the flavor mixing
 is generated by the anomaly term.  To clarify the structure of ${\bf Q}$,
 we note that the propagator of the scalar mesons in this model is written
 as follows;
\beq
{\bf D}_{\sigma}(q^2)= -[{\bf 1} + {\bf V}_{\sigma}\cdot {\bf \Pi}^{S}(q^2)]
^{-1} \cdot  {\bf V}_{\sigma},
\eeq
accordingly,
\beq
{\bf Q} = - {\bf D}_{\sigma}(0)\cdot {\bf V}_{\sigma}^{-1}.
\eeq
We remark that  when the interaction is absent
${\bf Q}= {\bf 1}\equiv {\bf Q}^{(0)}$.

As we have already noticed,
 it is more adequate to call ${\bf Q}$ the scalar charge (matrix)
of the  constituent quarks.
 Effective charges are usually enhanced (suppressed) due to
 collective excitations generated by the attractive (repulsive)
 forces.  In our case, we should have an enhancement because the
 interaction is attractive. The physical origin of the enhancement may
be described as  follows;
 the  external probe
 can interact  not only with the bare particle but also with the
 $q$-$\bar{q}$ cloud surrounding the bare particle.
 In our model,  the cloud is nicely
 summarized  by the scalar meson in the ring approximation.
  Such an enhancement (suppression) of a charge is
 well known in the many-body theory, especially in nuclear
physics\cite{BM}:
 The  effective electric charges of a nucleon in heavy nuclei is
enhanced due to the existence of the quadrupole giant resonances, while
 the effective axial charge $g_A$ is suppressed due to the
 existence of the Gamow-Teller giant resonance.
The sigma meson plays an analogous role to the
giant resonances in nuclei.

 It is also to be noted that the anomalous quark contents (the
degree of the violation of the OZI rule) are related
 with the mixing property of the scalar mesons.

The typical value for $Q_{ji}$ in the  parameter set  given in
 reads
\beq
{\bf Q} = \left ( \begin{array}{ccc}
1 + 1.14 & 0.45 & 0.15 \\
0.45 & 1 + 1.14 & 0.15 \\
0.28 & 0.28 & 1 + 0.42 \end{array}\right ),
\eeq
which shows the following:
(i)  The  nonperturbative  correlation in this channel
significantly increases the effective
 charge.
(ii)
The violation of the  OZI rule  is not negligibly small in this channel,
   as is seen from the off-diagonal elements.
(iii)   The $s$-quark mixing in the $u$ quark
 is small  in comparison  with  the $d$-quark mixing
 in $u$ ($Q_{us}/Q_{ud}$=0.15/0.45=1/3); this is  due to the mass
difference of $M_s$ and $M_d$.

Before using the full mass formula Eq. (3.3), let us see the
 result in the simple  additive quark ansatz, i.e.,
\beq
M_P\simeq 2 M_u + M_d,
\eeq
 for the proton, for example.  In this approximation, we have\cite{HK3b}
\beq
\Sigma_{\pi N} & = & 3\hat{m}(Q_{uu} + Q_{ud})= 43\ {\rm MeV},
\eeq
 where the $SU(2)$ isospin symmetry is assumed.
  If there were no interaction
 between current quarks, only the kinematical piece
 remains and gives $\Sigma_{\pi N} \simeq 3 \hat{m} \simeq (15- 24) $
MeV provided that $\hat m\simeq (5 - 8)$ MeV.  Thus one sees that
 the nonperturbative effect of the interaction responsible for
 the breaking of the chiral symmetry and the generation
 the dynamical mass is  necessary   to
 reproduce the empirical value $\Sigma_{\pi N} = (45 \pm 10) $ MeV.
We should also stress again  that the nonperturbative effect is intimately
 related to the existence of the collective excitation in this
channel, i.e., the sigma meson.

The strangeness content of the proton is  written in the same approximation
as
\beq
\la \bar{s}s \ra_p = 3 {{\partial M_u} \over {\partial m_s} }
= 3 {{\partial M^D_u} \over {\partial m_s} },
\eeq
which shows that the OZI-violating matrix element has the
 nonperturbative origin.
Numerically, the ratio
$y \equiv 2{\la  \bar{s}s \ra }_N / {\la  \bar{u}u + \bar{d}d \ra }_N $
is as small as 0.12.
 Thus we have seen that the large empirical value of $\Sigpi$
 can be obtained without a large strangeness content of the
nucleon.\cite{HK3b}

How about the effect of the residual spin-spin interaction and the
confinement included in the mass formula Eq.(3.3).\cite{HK3d}
 Using the mass formula instead of the additive ansatz, we have
 for the proton, for example,
\beq
 \langle \bar uu\rangle_P = 4.97 \ (4.73), \ \
  \langle \bar dd \rangle_P = 4.00 \ (3.03),
 \langle \bar{s}s \rangle_P = .53 \ (0.46),
\eeq
hence
\beq
\Sigpi = 49 {\rm MeV}.
\eeq
Here the numbers in the paranthesis are the results in the
additive ansatz.  One sees that the residual interactions
enhance the $\Sigpi$ and the empirical value is reproduced.

Here we should mention the cutoff-scheme dependence on $\Sigma_{\pi N}$
\cite{HK3f}.
The canonical (three momentum) cutoff
  is the most reasonable scheme from the physical point of
view \cite{HK94}
 and  Eq.(3.15) is in fact calculated in  this scheme.
 In the  other schemes \cite{BJM},
  smaller  $\Sigma_{\pi N}$ are obtained.
  This  is due to
 the  unphysical quenching of $\la \bar{q}q \ra$ (or $M$)
 for large  $m$, and   such a quenching gives
 small $Q_{ij}$ and thus small $\Sigma_{\pi N}$.

\subsection{The convergence radius of the chiral perturbation
 theory and the mass of the sigma meson}

In the chiral perturbation theory\cite{LiPa}, observables are expanded
 with respect to  the current quark masses.
The general structure of such an  expansion is known to be
\beq
{\cal O}(m) = {\cal O}(0) + \sum_{l=1}^{\infty} a_l\  m^l + b_1\, m^{1/2}
 + b_2\, m \ln m + \cdot \cdot \cdot \ \ .
\eeq
Here the first term denotes the value at the chiral limit,
 the second term is analytic terms and the last terms are
 non-analytic ones.  The non-analyticity is  due to the masslessness
 of the NG-boson in the chiral limit, while
 the coefficients of the analytic terms ($a_i$) depend on the details
of the QCD dynamics.
 Carruthers and Haymaker (CH) examined  the convergence property
of this analytic series \cite{CH},
 using the $SU(3)$ linear $\sigma$-model.  They found that
 the convergence radius of the analytic series
 is  so small that one
  cannot reach the physical region of the current quark masses
 for the strangeness sector.
 Hatsuda \cite{HK3e} examined the same problem using the NJL model and
 reached a similar conclusion with CH.

In this subsection, we shall indicate that the convergence radius is
intimately related with the mass of the sigma meson in the chiral limit.
The following is  mostly based on section 3.5 of ref. \cite{HK3e,HK94}.

For simplicity, let us
take the effective potential of the linear $\sigma$ model in the tree
level  with an explicit symmetry breaking,
\begin{eqnarray}
V(\sigma) & = & V_0(\sigma)  - f_{\pi}m_{NG}^2
  \sigma\ \ ,\quad \  \ V_0(\sigma)=
 {m_{\sigma}^2 \over 8 f_{\pi}^2}(\sigma^2
- f_{\pi}^2)^2\ \ .
\end{eqnarray}
$m_{\sigma}$ denotes  the second derivative of $V_0(\sigma)$
with respect to $\sigma$ and is identified as
the mass of  the scalar meson in the chiral limit.
  $m_{\sigma}$ sets the  characteristic scale of the system.
$m_{NG}$ is the mass of the NG boson which vanishes in the chiral
limit.
   The vacuum is determined as the point where the effective potential
takes the minimum;
$d V(\sigma) / d \sigma =0,$
 or equivalently
${ d V_0/ d \sigma } = f_{\pi} m_{NG}^2 ,$
which is reduced to
\begin{eqnarray}
F(x)\equiv x (x^2-1) = 2 ({ m_{\pi} \over m_{\sigma}})^2
  \equiv a \ \ \ \ {\rm with}   \ \ \  x=\sigma/f_{\pi}.
\end{eqnarray}
$ V_0(\sigma)$ and $F(x)$ are shown in Fig.3.6 of ref. \cite{HK94}.
These equations  actually determine the extrema of the effective
potential.
 $F(x)$ represents how the effective potential is steep at $x$.
One sees that  $a$ is a natural dimensionless expansion parameter.
Notice that $a$ is the ratio of pion mass squared to the sigma meson
mass squared.   In the chiral limit ($a=0$), Eq.(3.22) has
three solutions $x=0$ (the symmetric phase) and $x=\pm 1$ (the
dynamically broken phases).  For $a\neq 0$,
 the attempt to calculate $x(a)$ by the perturbation around $x(a=0)=1$ fails
at the point where $dx(a)/da=\infty $.  This condition gives the
 convergence radius
\beq
  \mid a \mid < a_{cr} = 2/3\sqrt{3}.
\eeq
Actually, the existence of $a_{cr}$  is related to the behavior of
$x(a)$ for $a <0$:
 Note that  the convergence radius of the
 Taylor series $x(a)= \sum_n c_n a^n$ is determined  by the nearest
singularity of $x(a)$ in the complex $a$-plain.
 For  $a$ smaller than $-2/3\sqrt{3}$, there are no  solutions
which continuously connected to the solution $x=1$ in the chiral
limit. We remark that the critical value $a_{cr}$ is the maximum value of
 the steepness of the effective potential  in the chiral limit.
In the infinite steepness limit,  the sigma meson mass goes infinity and
 the linear sigma model is reduced to a non-linear sigma model.
 Therefore, it is natural that the convergence radius in the physical
unit is related to the sigma meson mass, as we will see now.

Let's translate $a_{cr}$ into the critical value of $m_{NG}$:
Owing to (3.23), $m_{NG}^2$ should satisfy the following relation
 \beq
m_{NG}^2 < {m_{\sigma}^2 \over 3 \sqrt{3}}.
\eeq
In the NJL model,
  $m_{\sigma} \simeq 2M_u \simeq 700$ MeV, as noted previously.
 Weinberg \cite{WEIN} suggests
 Re $m_{\sigma} \simeq$ Re $m_{\rho}\simeq$ 770 MeV on the basis of
 the mended symmetry.
 If the existence of the sigma meson with such a rather small mass
 is confirmed, we have $ \sim(340 {\rm MeV})^2$ for the right hand
side of Eq. (3.24). It means that
 $m_{\pi}^2=(140 {\rm MeV})^2$  lies within the convergence radius,
while $m_{K}^2 = (495 {\rm MeV})^2$ does not, implying
  that the expansion of observables in the strange sector
with respect to  $m_{K}$  is doubtful.
 Conversely, for $m_K$ to lie within the convergence radius,
 $m_{\sigma}$ in the chiral limit have to  be as large as 1.1 GeV.

In summary, we have seen that the convergence radius is related to
 how steep the effective potential is in the chiral limit. Since the
 steepness can  be translated to the sigma meson mass, the convergence
radius may be  related to the properties  of the sigma meson. The NJL
model and the mended symmetry as well as the linear sigma model
 suggest that the strangeness sector may be dangerous to apply the
chiral perturbation theory.
 Nevertheless, the final answer will be given by
  model-independent analyses of the effective potential or the
nature of the chiral symmetry breaking in QCD.  With such  analyses,
 one could also have a good insight into the sigma  meson.

%\newpage
\setcounter{section}{3}
\setcounter{equation}{0}

\section{Producing and detecting the sigma meson in experiment}

\setcounter{equation}{0}
\renewcommand{\theequation}{\arabic{section}.\arabic{equation}}

We  have seen that the correlations which may be summarized by the
 unstable and hence elusive sigma meson play significant roles in the
 hadron phenomenology at low energies.  Therefore one may wonder
 whether  there is any chance   to  observe the sigma meson clearly.
 What does come when the environment is changed by rising  temperature
 and/or density?
 As was  first shown by Hatsuda and the present author\cite{HK1f,HK2a},
 the sigma meson decreases
 the mass in association with the chiral restoration in the hot and/or
dense medium, and  the width of the meson is also expected to decrease
 because the pion hardly changes the mass as long as the system is in
 the Nambu-Goldstone phase.
 Thus one can  expect a  chance to see the sigma meson as a sharp
 resonance at high temperature.
Such a behavior of the meson may be detected by observing two pion with
 the invariant mass around several hundred MeV in relativistic heavy
ion collisions.  As was indicated by Weldon\cite{WELDON},
 when the charge pions have
 finite chemical potentials, the process $\sigma \rightarrow \gamma
 \rightarrow $2 leptons can be used to detect the sigma meson.

 It is worth mentioning that
 the simulations of the lattice QCD \cite{GOTT} show the decrease of the
 screening
mass $m^{\sigma }_{\rm scr}$ of the sigma meson. Here a screening mass is
 defined through the space correlation of the relevant operator rather
 than the time correlation as a dynamical (real) mass.
The relation between  the screening mass and the dynamical mass as
discussed by  Hatsuda and Kunihiro is not clearly understood yet.
 Nevertheless it is known \cite{FRIMAN} that the NJL model
gives the similar behavior for the screening masses in the scalar
channels with the dynamical ones. It means that the lattice result
 on the screening masses may suggest that the dynamical masses also
 behave in a way as predicted in \cite{HK2a}.
\footnote{
It is interesting that the decrease of the scalar meson in the medium is also
 shown \cite{saito} with  hadron models like
the $\sigma$-$\omega$ model\cite{WALECKA},
although the role of the chiral symmetry is obscure in such a
 phenomenological model.}

We remark that the logic of this type of physics is
not new for nuclear physicists.  Studying a possible change of
collective modes being associated with a change of the ground state of
 nuclei is in fact what they have usually been doing: When nuclei are
 being deformed from a spherical shape, the vibrational modes will
 soften, i.e., the energy of the 2$^{+}$\  phonon is decreased.
 If a nucleus is near the superconducting phase, the pairing
vibrational mode is also softened.\cite{BM} For more than a decade,
  people have been eager to  search a precursory collective
phenomena for the pion condensation in nuclei.\cite{toki}
These are all the same physics in the logic.
 In fact the analogy of  phase transitions in  many-body systems with
the chiral transition in QCD is emphasized in \cite{HK1d}.

Here we  propose several experiments for examining possible
restoration of chiral  symmetry in nuclei and the possible existence of
the sigma meson. One uses pions, another protons and light nuclei and
the other electrons. To detect the sigma, one may use 4 $\gamma$'s
and/or two leptons. The latter
 process is  possible because a scalar particle can be converted to
 a vector particle because of the scalar-vector mixing in the
 system with   a finite baryonic density. This mixing  is well known
in the  Walecka  model\cite{WALECKA}. Microscopically, the process is
described by
  $\sigma \rightarrow {\rm N} \bar {\rm N} \rightarrow \gamma$.
  Here $\sigma $ may be replaced by any scalar particle, and
 $\gamma$ any vector particle with the same quantum numbers other than spin
 and parity.

\begin{flushleft}
{\bf 1.\ A\ ($\pi$, 4$\gamma$\  N)\ A$'$}
\end{flushleft}

%\begin{minipage}[t]{6.5cm}
This reaction process is depicted in Fig. 3.
 In this reaction, the charged pion ($\pi ^{\pm}$) is absorbed by a nucleon in
 the nucleus, then the nucleon emits the sigma meson, which decays into
 two pions.  To make a veto for the two pions from the rho meson, the produced
 pions should be neutral ones which decay to four $\gamma$ 's.

\vspace{.5cm}
%\end{minipage}
%\begin{flushright}
%\begin{minipage}[thb]{6.7cm}

{\cl {\bf \fbox{Fig.3}}}
%\vspace{1.cm}
%\end{minipage}
%\end{flushright}

\vspace{.5cm}
\begin{flushleft}
{\bf 2. \ A\ (P, 4$\gamma$\  N)\ A$'$}
\end{flushleft}

%\begin{minipage}[t]{6.8cm}
This type of the reaction is depicted in Fig. 4.  The incident proton,
 deuteron or $^3$He ... collides with a nucleon in the nucleus, then the
 incident particle will emit the sigma meson, which decays into two pions.
 One may detect 4 $\gamma$ 's from 2 $\pi ^0$.  The
 collision  with a nucleon may occur after  the emission of the sigma meson;
 the  collision process is needed for the energy-momentum matching.

\vspace{.5cm}
{\cl {\bf \fbox{Fig.4}}}
\vspace{.5cm}

In the detection, one may use the two leptons from the process
 $\sigma \rightarrow {\rm N} \bar {\rm N} \rightarrow \gamma$ mentioned above:
  See Fig. 5.
 This detection may gives a clean data, but the yield might be small.

\vspace{.5cm}
{\cl {\bf \fbox{Fig.5}}}
\vspace{.5cm}

\begin{flushleft}
 {\bf 3. \  A\ (e$^{-}$, 4$\gamma$\   e$^{-}$)\ A$'$}
\end{flushleft}

%\begin{minipage}[t]{6.5cm}
 The final example uses the electron beam; See Fig. 6.
 The $\gamma$ ray emitted from the electron is  converted to the omega
 meson in accord with the vector meson dominance principle. The omega meson
 may decay into the sigma meson in the baryonic medium via the process
 $\omega \rightarrow $ N $\bar{\rm N} \rightarrow \sigma$. The sigma will
 decay into two pions. One may detect the 4$ \gamma$'s from the 2 neutral
 pions.

\vspace{1cm}
{\cl {\bf \fbox{Fig.6}}}
\vspace{1cm}

%\newpage
\setcounter{section}{4}
\setcounter{equation}{0}
\section{Fluctuation effects in hot quark matter and superconductor}
\setcounter{equation}{0}
\renewcommand{\theequation}{\arabic{section}.\arabic{equation}}

A decade ago, in the paper \cite{HK85} entitled ``Fluctuation effects in
Hot Quark Matter; Precursors of Chiral Transition at Finite Temperature",
 Hatsuda and the present author showed that there should  exist
 colorless hadronic  modes even in the deconfined and chirally symmetric
phase at high temperature, if the chiral transition is not of
 so strong first order. They are actually the fluctuations of the
order parameter $\lla (\bar {q} q)^2\rra$ and $\lla (\bar {q}i\gamma_5
\tau q)^2\rra$. The excitation energies and the widths
of the modes decrease  as the system approaches the critical point from
 the above. Thus the modes are a kind of soft modes.

Such  long-range fluctuations near the critical point are
 actually  well-known phenomena in many body systems. Apart from the
 modes in nuclei  mentioned in the preceding section,
some examples are the soft phonons in
 (anti-)ferroelectrics, the paramagnon in ferromagnets,
 the paring vibration in nuclei \cite{HK94}.

In subsequent papers \cite{HK2h},
the present author extended the discussion  to the vector channel:
 The quark number susceptibility is related to the
 fluctuation in the vector channel, and the recent lattice simulations
\cite{QNS}  can be interpreted as showing that the interaction in the vector
 channels  is  weak at high temperatures.

Now, it is well known that the NJL model was invented on the basis of an
analogy between the superconductor and the chirally broken phase in the
 vacuum\cite{NJL}.  Therefore it is interesting that there seems no
 propagating
 soft modes in the material which undergoes the phase transition into the
 superconducting phase.  Actually, dynamic critical phenomena in
 superconductors can be
well described by the time-dependent Ginzburg-Landau equation  which is
 a diffusion equation\cite{TDGL}.  It has been recently shown \cite{Kuni94}
 that   this is due to the
fact that the number of the effective spatial dimensions of
the superconductor in the BCS  model is one:  In fact, the excitation
 energy for the precursory mode in the BCS model is given by
\cite{kadanof}
\beq
1 + g_{BCS} \int_{-\omega_D}^{\omega_D}d\epsilon
\frac{\tanh \beta \epsilon/2}{\epsilon - \omega/2}=0,
\eeq
where $g_{BCS}$ is a constant proportional to the coupling constant and
 $\omega_D$ is the Debye frequency,
while the corresponding equation in the NJL model is given by
\beq
1 + g_{NJL} \int_{0}^{\Lambda}k^2dk
\frac{\tanh \beta k/2}{k - \omega/2}=0,
\eeq
with $g_{NJL}$ being a constant proportional to the coupling constant.
These equations show that the BCS model describe the system essentially
 a one-dimensional system because the cutoff $\omega_D$ is small; this
means that the frequency region where the electrons feel the attraction
 is narrow.  Thus the precursory mode in the superconductor does not become
 a propagating one, and the equation describing the low-frequency phenomena
 is well given by the diffusive TDGL equation\cite{TDGL}.
  This is in contrast to the
 chiral transition as described by the NJL model, where the precursory mode
 is propagating\cite{HK85}; for the details, see ref.\cite{Kuni94}

%\vspace{.5cm}{\cl {\fbox{\bf Fig.7}}}\vspace{.5cm}

%\newpage
\setcounter{section}{5}
\setcounter{equation}{0}

\section{The saturation property of nuclei and chiral symmetry}

\setcounter{equation}{0}
\renewcommand{\theequation}{\arabic{section}.\arabic{equation}}

We have shown  and stressed that most of the low-energy phenomena in
QCD are largely determined by chiral symmetry and its dynamical breaking
 rather than the confinement\cite{HK94}; an interplay between the
explicit breaking of chiral symmetry due to the current quark masses and the
$U_A(1)$ anomaly adds some interesting variations to the low-energy
hadron dynamics.  Then how about nuclei, stable many-hadron systems.
 Actually, they  are only the bound systems of hadrons, apart from the
 possible H-dibaryon conjectured by Jaffe\cite{jaffeh}.
Here, we would like to indicate that the deepest reason of the
 stability of nuclei is the chiral symmetry and its dynamical
 breaking in QCD.

The stability of the nuclei can be summarized by the saturation properties
of the binding energy and density\cite{bethe}:
All nuclei except for some smallest nuclei have the almost
 the same binding energy $E_B$ per nucleon  and the central density
 $\rho$ irrespective of the nucleon number.
  The corresponding values of the nuclear matter are
\beq
E_B\simeq -15 \ \ {\rm MeV}, \ \ \ \ \rho = .17 \ \ {\rm fm}^3\equiv \rho_0
 \ \ (k_F= 1.36 {\rm fm}^{-1}).
\eeq
One may notice that

\noindent
(i) $E_B$ is small compared with the rest mass $\sim 1$ GeV; nuclei are
 loosely bound system.

\noindent
(ii) the nuclear density $\rho_0$ is remarkably low; it
 means that the inter-nucleon distance is as distant as 1.8 fm,
 which is much larger than the radius ($\sim 0.4$ fm) of the repulsive core
of the nuclear force.

Why?
The answer was given by
the nuclear matter theory based on the reaction matrix (G-matrix)
initiated by Brueckner and developed in  the 60's.\cite{bethe}

The resultant nuclear density is so low that the nuclear matter may be
 treated as a good approximation by the superposition of the two-nucleon
 interactions, i.e., the independent-pair approximation.
The effective two-nucleon potential (G-matrix) is
then constructed by a ladder approximation which takes into account
the Pauli principle in the intermediate states.
The reaction matrix is expanded by partial waves with specific relative
 angular momenta.  The main attraction comes from the $S$-waves, $^1S_0$
 and $^3S_1$; the former is primarily due to the iso-singlet scalar
meson in the one-boson-exchange (OBE) model, while the latter due to the
 pion.  The former contribution in the absolute value only increases with
 the density.  The  $^3S_1$-wave contribution $V_{eff}^{^3S_1}(\rho)$
actually has  a coupling to the $^3D_1$ wave and shows
 a peculiar density dependence as shown in Fig.7;
$V_{eff}^{^3S_1}(\rho)$ in the absolute value takes the minimum about
 the density $\sim \rho_0$, and then the attraction becomes smaller.
 This is the primary reason of the saturation properties of the
 nuclear matter\cite{bethe}.  The peculiar density dependence is due to the
coupling between the $^3S_1$ and the $^3D_1$ waves by the tensor force
 of the one-pion-exchange potential (OPEP); the Pauli principle in the
 intermediate states in the coupling diagrams gives the peculiar
 bending behavior.

Now the tensor force of OPEP  arises because
 the pion is pseudo-scalar. The density of the bending point is so low
 because the pion is light.   Needless to say, the pion
 is pseudo-scalar and light because it is the NG-boson of the $SU(2)
\otimes SU(2)$ chiral symmetry.
Thus one may say the saturation properties of the
 nuclear matter and hence the stable existence of nuclei are due to the
 chiral symmetry and its dynamical breaking.  A few remars are in order:
As for the determination of the
 range of the OPEP  or the tensor force of the nuclear force, the
 subtle cancellation between the OPEP and the potential due to the
 $\rho$ meson exchange in the tensor coupling is important.  It is
 also to be remarked  that the actual value  of the pion mass
is determined by the
 small current quark masses of u and d quarks. Therefore
 if they were much larger
 or smaller, nuclei might have not existed or the universe would
 be  much different from the one where we exist.

\vspace{.5cm}
{\cl {\fbox{\bf Fig.7}}}

\vspace{3cm}

{\cl {\bf {\large Acknowledgement}}}

\vspace{0.5cm}

A part of this report is based on the works summarized in \cite{HK94} done
in collaboration with Tetsuo Hatsuda; the present author thanks
 him  for his collaboration.
The present author also expresses his sincere thanks to Professor R. Tamagaki,
who guided him as a supervisor in his graduate school to the
 nuclear matter theory as given in \cite{bethe} and to the physics of
 high-density nuclear matter.  He is also grateful to
 Professor T. Tsuneto for discussions on time-dependent phenomena in
 superconductors. Hajime Shimizu is also gratefully
 acknowledged for discussions on
 possible experiments to produce the sigma meson in a nucleus.
 This work was supported by the Japanese Grant-in-Aid for Science Research
 Fund of the Ministry of Education, Science and Culture, No. 05804014 and
 Joint Research Center for Science and Technology, Ryukoku University.

%\newpage

%\newpage
{\bf Figure Captions}

\begin{description}
\item[Fig.1]
The data and the calculated phase shift of pion-pion scattering in
 $I=J=0$ channel.\cite{kamin}

\item[Fig.2]
  The diagram which enhances the process with $\Delta I =1/2.$

\item[Fig.3]
 The diagram describing the pion-production of the sigma meson.
 The sigma meson may be detected in a bump of the invariant mass of the
 4 $\gamma$.

\item[Fig. 4]
 The diagram describing the nucleon- or light nucleus-production
  of the sigma meson.

\item[Fig.5]
 The diagram describing the conversion of the sigma meson to
 the gamma ray.

\item[Fig. 6]
 The diagram describing the electro-production  of
the sigma meson.

%\item[Fig. 7]

\item[Fig. 7]
Contribution to the binding energy of nuclear matter of the $^1S$
 and $^3S$ state as a function the Fermi momentum $k_F$. The calculation is
 by D.\ W.\ L.\  Sprung. Taken from a review by Bethe \cite{bethe}.

\end{description}

\end{document}